\documentclass[Physsubmission, Phys]{SciPost}

\usepackage{amsmath}
\usepackage{graphicx}
\usepackage{dcolumn}
\usepackage{bm}
\renewcommand{\labelenumi}{(\roman{enumi})}
\usepackage{comment}
\usepackage{braket}
\usepackage{appendix}
\usepackage{physics}
\usepackage{mathtools}
\usepackage{empheq}
\usepackage{pifont}

  \newcommand{\xmark}{\color{red}\ding{55}}%
  \newcommand{\gcheckmark}{\color{green}\ding{51}}%

\hypersetup{
    colorlinks,
    linkcolor={red!50!black},
    citecolor={blue!50!black},
    urlcolor={blue!80!black}
}

\usepackage[bitstream-charter]{mathdesign}
\DeclareSymbolFont{usualmathcal}{OMS}{cmsy}{m}{n}
\DeclareSymbolFontAlphabet{\mathcal}{usualmathcal}

\begin{document}

\begin{center}{\Large \textbf{
Rigorous reconstruction of gluon propagator in the presence of complex singularities\\
}}\end{center}

\begin{center}
Yui Hayashi\textsuperscript{1$\star$} and
Kei-Ichi Kondo\textsuperscript{1,2}
\end{center}

\begin{center}
{\bf 1} Department of Physics, Graduate School of Science and Engineering, Chiba University, Chiba 263-8522, Japan
\\
{\bf 2} Department of Physics, Graduate School of Science, Chiba University, Chiba 263-8522, Japan
\\
* yhayashi@chiba-u.jp
\end{center}

\begin{center}
\today
\end{center}


\definecolor{palegray}{gray}{0.95}
\begin{center}
\colorbox{palegray}{
  \begin{minipage}{0.95\textwidth}
    \begin{center}
    {\it  XXXIII International (ONLINE) Workshop on High Energy Physics \\“Hard Problems of Hadron Physics:  Non-Perturbative QCD \& Related Quests”}\\
    {\it November 8-12, 2021} \\
    \doi{10.21468/SciPostPhysProc.?}\\
    \end{center}
  \end{minipage}
}
\end{center}

\section*{Abstract}
{\bf
It has been suggested that the Landau-gauge gluon propagator has complex singularities, which invalidates the Källén–Lehmann spectral representation. Since such singularities are beyond the standard formalism of quantum field theory, the reconstruction of Minkowski propagators from Euclidean propagators has to be carefully examined for their interpretation. In this talk, we present rigorous results on this reconstruction in the presence of complex singularities. As a result, the analytically continued Wightman function is holomorphic in the usual tube, and the Lorentz symmetry and locality are kept valid. On the other hand, the Wightman function on the Minkowski spacetime is a non-tempered distribution and violates the positivity condition. Finally, we discuss an interpretation and implications of complex singularities in quantum theories, arguing that complex singularities correspond to zero-norm confined states.
}

\vspace{10pt}
\noindent\rule{\textwidth}{1pt}
\tableofcontents\thispagestyle{fancy}
\noindent\rule{\textwidth}{1pt}
\vspace{10pt}

\section{Introduction}

Correlation functions are essential building blocks of a quantum field theory (QFT), and their analytic structures provide an insight into the state space. 
In the last decades, correlation functions in the Landau gauge have been studied by both lattice and continuum methods to understand fundamental aspects of quantum chromodynamics (QCD) as well as hadron phenomenology.

In particular, two-point functions, or propagators, have important information on QFT.
For example, the K\"all\'en--Lehmann spectral representation implies the correspondence between singularities of a propagator $D(k^2)$ and states $\ket{P_n}$ non-orthogonal to the state $\phi(0) \ket{0}$:
\begin{align}
D(k^2) &= \int_0 ^\infty d \sigma^2 \frac{\rho(\sigma^2)}{\sigma^2 - k^2}, \label{eq:KL}\\
   \theta(k_0) \rho(k^2) 
&:=  (2\pi)^{3} \sum_{n }  |\langle 0 | \phi(0) | P_n \rangle|^2 \delta^4(P_n-k).
\end{align}
Observing an analytic structure would give a valuable hint for understanding fundamental aspects of QCD, for example, the color confinement.

Therefore, based on the progress in the QCD correlation functions, there has been an increasing interest in analytic structures of the QCD propagators in recent years.
Some results of recent independent approaches, e.g., numerical reconstruction techniques from Euclidean data \cite{BT2019, Falcao:2020vyr}, models of massivelike gluons \cite{Siringo16, HK2018, Hayashi:2020few}, and the ray technique of the Dyson-Schwinger equation \cite{SFK12,Fischer-Huber}, suggest that the Landau-gauge gluon propagator has \textit{complex singularities}, which are unusual singularities invalidating the K\"all\'en-Lehmann spectral representation.

On the other hand, implications of complex singularities have been less studied. There are only old works \cite{Stingl} discussing this subject heuristically.
However, since complex singularities are beyond the standard formalism of QFT, we need to consider their interpretation carefully.
Hence, we study the rigorous reconstruction of propagators with such singularities \cite{Hayashi:2021nnj, Hayashi:2021jju}.

In this presentation, we sketch out the reconstruction of propagators and its consequences in the presence of complex singularities.

\section{Definition and main questions}

We point out that complex singularities are defined in terms of Euclidean propagators.
Therefore, the reconstruction procedure from Euclidean field theory to quantum field theory should be carefully considered.
We then pose the main questions addressed in this presentation.

\subsection{Definition of complex singularity}

For starting a rigorous discussion, an appropriate definition should be provided.

\begin{figure}[t]
    \centering
    \includegraphics[width =  0.7 \linewidth]{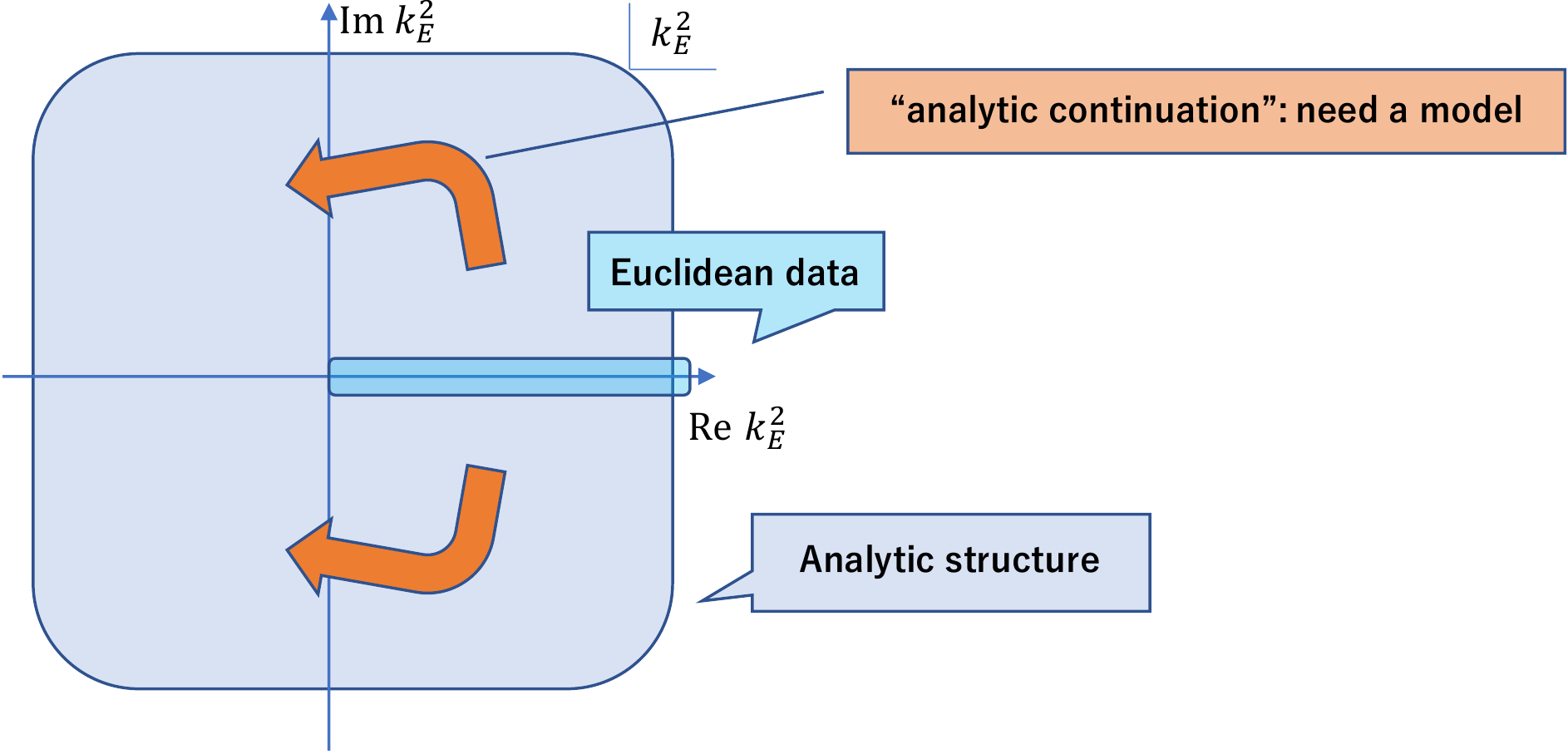}
    \caption{Conceptual picture describing methodology of how an analytic structure is investigated in the literature. Note that what we examine here is a structure on the complexified Euclidean momentum plane.}
    \label{fig:concept_analytic}
\end{figure}

We begin by reviewing how the analytic structures are investigated in the literature.
Roughly speaking, an analytic structure is obtained by an ``analytic continuation'' from Euclidean data (Fig.~\ref{fig:concept_analytic}).
Obviously, there exists a fundamental issue; an analytic continuation from finite data is not unique.
The best thing we can do is a speculative study of an analytic structure using a model.
If we have a model with some theoretical backgrounds, the model propagators can provide possible analytic structures of the QCD propagators.
In this way, the analytic structures have been examined.

We emphasize that the analytic structure to be obtained is that of an analytically-continued \textit{Euclidean} propagator.
Therefore, we define complex singularity as \textit{singularity off the real axis in the complex momentum $k_E^2$-plane of an analytically-continued Euclidean propagator}.

For technical reasons, we further assume the following properties for complex singularities: (1) boundedness of complex singularities in $|k_E^2|$, (2) holomorphy of $D(k_E^2)$ in a neighborhood of the real axis except for the timelike ($k_E^2 < 0$) singularities, (3) some regularity of the timelike singularities.

\subsection{Main questions}

Since complex singularity is a property of the Euclidean propagator, we need a reconstruction to obtain its interpretation.

\begin{figure}[t]
    \centering
    \includegraphics[width =  0.7 \linewidth]{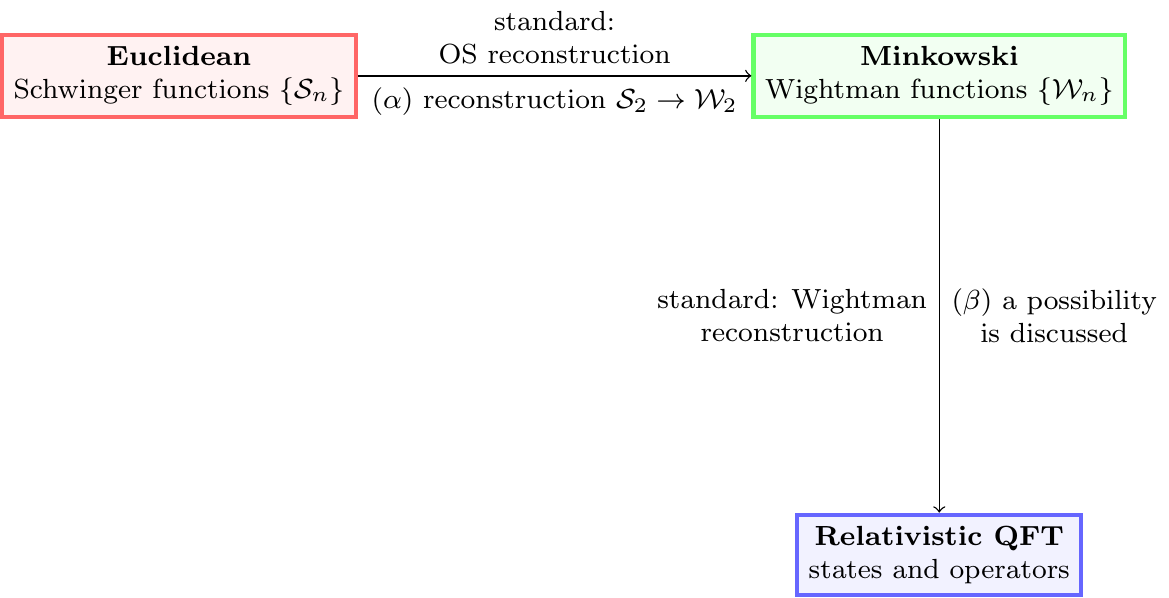}
    \caption{Standard reconstruction procedure and contents of our study ($\alpha$) and ($\beta$). Taken from \cite{Hayashi:2021nnj}.}
    \label{fig:reconstruction}
\end{figure}

To clarify what we should address, let us briefly summarize how we reconstruct quantum theories from Euclidean field theories in the standard formalism \cite{OS73, OS75} (Fig.~\ref{fig:reconstruction}).
We start with a set of Euclidean correlation functions, called Schwinger functions.
If these Schwinger functions satisfy the Osterwalder-Schrader (OS) axioms, we can reconstruct the Wightman functions on the Minkowski spacetime by an analytic continuation, which satisfy the Wightman axioms.
Subsequently, by the Wightman reconstruction, we can obtain a quantum theory written in terms of states and operators from the Wightman functions.

The natural question here is whether or not we can do the same thing in the presence of complex singularities.
In what follows, we mainly discuss the following two questions corresponding to the arrows ($\alpha$) and ($\beta$) in Fig.~\ref{fig:reconstruction}.

\begin{itemize}
    \item [($\alpha$)] Is it possible to reconstruct a Wightman function $W(\xi^0, \vec{\xi})$ on the Minkowski spacetime from the Schwinger function? Which conditions of the Wightman/OS axioms are preserved/violated?
    \item [($\beta$)] Does there exist a quantum theory reproducing the reconstructed Wightman function $W(\xi^0, \vec{\xi})$ as a vacuum expectation value: $W(\xi) = \langle 0 | \phi(\xi) \phi(0) | 0 \rangle$?
    If it exists, what states cause complex singularities?
\end{itemize}

We will answer these questions affirmatively \cite{Hayashi:2021jju, Hayashi:2021nnj}.

\section{Reconstruction of the Wightman function and its general properties}

Let us move on to the first topic $(\alpha)$.
We reconstruct the Wightman function $W(t,\Vec{x})$ from the Schwinger function with complex singularities by identifying the Schwinger function as imaginary-time data of the Wightman function: $S(\tau, \Vec{x}) = W(- i \tau ,\Vec{x}) ~~(\tau > 0)$.

To answer the question $(\alpha)$, we proved \cite{Hayashi:2021jju, Hayashi:2021nnj}:
\begin{enumerate}
\renewcommand{\labelenumi}{(\Alph{enumi})}
    \item The reflection positivity
    is violated for the Schwinger function.
    \item The holomorphy of the Wightman function $W(\xi - i \eta)$ in the tube $\mathbb{R}^4 - i V_+$ and the existence of the boundary value as a distribution are still valid, where $V_+$ denotes the (open) forward light cone. Thus, we can reconstruct the Wightman function from the Schwinger function.
    \item The temperedness and the positivity condition are violated for the reconstructed Wightman function. The spectral condition is never satisfied since it requires the temperedness as a prerequisite.
    \item The Lorentz symmetry and spacelike commutativity are kept intact.
\end{enumerate}

Let us see these properties with a simple example: one pair of complex conjugate poles (e.g., the typical Gribov-Zwanziger fit),
\begin{align}
    D(k_E^2) = \frac{Z}{k_E^2 + M^2} + \frac{Z^*}{k_E^2 + (M^*)^2}.
\end{align}
Since any complex singularity can be written as a ``sum'' of complex poles from the Cauchy integration formula, this example will capture the essential features of complex singularities.
For detailed proofs of these results, see \cite{Hayashi:2021nnj}.

The Schwinger function in the position space reads
\begin{align}
    S (\vec{\xi}, \xi_4) = \int \frac{d^3 \vec{k}}{(2 \pi)^3} e^{i\vec{k} \cdot \vec{\xi}} \left[ \frac{Z}{2 E_{\vec{k}} } e^{- E_{\vec{k}} |\xi_4|} +  \frac{Z^*}{2 E_{\vec{k}}^*} e^{- E_{\vec{k}}^* |\xi_4|}
    \right], \label{eq:simple_complex_poles_Schwinger}
\end{align}
where $E_{\vec{k}} = \sqrt{\vec{k}^2 +M^2}$ is a branch of $\operatorname{Re} E_{\vec{k}} > 0$.

(B) 
We now analytically continue the Wightman function starting from the imaginary-time data $S(\vec{\xi} , \xi_4) = W(- i \xi_4, \vec{\xi})$.
The straightforward integral representation,
\begin{align}
    W&(\xi - i\eta) = \int \frac{d^3 \vec{k}}{(2 \pi)^3} e^{i\vec{k} \cdot (\vec{\xi} - i \vec{\eta})} \left[ \frac{Z}{2 E_{\vec{k}} } e^{- i E_{\vec{k}} (\xi^0 - i \eta^0)} +  \frac{Z^*}{2 E_{\vec{k}}^*} e^{- i E_{\vec{k}}^* (\xi^0 - i \eta^0)} 
    \right], \label{eq:simple_complex_poles_hol_Wightman}
\end{align}
provides a desired analytic continuation to the tube $\mathbb{R}^4 - iV_+$.
Indeed, this expression is holomorphic in the tube $\xi - i \eta \in \mathbb{R}^4 - iV_+$ since the integrand decreases rapidly in $|\vec{k}|$ for $\eta \in V_+$.

We can take the ``limit'' $\eta \rightarrow 0~ (\eta \in V_+)$ of (\ref{eq:simple_complex_poles_hol_Wightman}) as a distribution\footnote{A subtle point here is the integral over $\vec{k}$, which is just the Fourier transformation and can be defined properly as a distribution.}:
\begin{align}
    W&(\xi) = \int \frac{d^3 \vec{k}}{(2 \pi)^3} e^{i\vec{k} \cdot \vec{\xi} } \left[ \frac{Z}{2 E_{\vec{k}} } e^{- i E_{\vec{k}} \xi^0} +  \frac{Z^*}{2 E_{\vec{k}}^*} e^{- i E_{\vec{k}}^* \xi^0}
    \right]. \label{eq:Wightman}
\end{align}

(C) The Wightman function (\ref{eq:Wightman}) grows exponentially in $\xi^0$ since $E_{\vec{k}}$ is complex. Therefore, the Wightman function on the Minkowski spacetime violates the temperedness.

The violation of positivity can be proved by the nontemperedness.
For this, we show
\begin{align}
    \mathrm{(Positivity) \Rightarrow (Temperedness)}.
\end{align}
Intuitively, this can be understood as follows.
\begin{enumerate}
    \item The positivity of $W(\xi)$ corresponds to the positivity of the sector $\{ \phi(x) \ket{0} \}_{x \in \mathbb{R}^4}$.
    \item The translational invariance of the two-point function corresponds to the unitarity of the translation operator $U(a)$ defined on this sector: $U(a) \phi(x) \ket{0} := \phi(x + a) \ket{0}$.
\end{enumerate}
These observations lead to a ``upper bound'' on $|W(a)| = | \langle 0 | \phi(0) U(-a) \phi(0) | 0 \rangle |  \leq  | \langle 0 | \phi(0)  \phi(0) | 0 \rangle | $, which will imply that $W(a)$ is tempered\footnote{Of course, since $W(\xi)$ is a distribution, the upper bound does not exist. Nevertheless, we can also prove the claim rigorously in the same spirit.}.

(A) 
Similarly, the violation of the reflection positivity can be shown by the nontemperedness. By repeating a part of the Osterwalder-Schrader reconstruction \cite{OS73} from Schwinger functions to Wightman functions, the reflection positivity yields the temperedness of the Wightman function.

For the example (\ref{eq:simple_complex_poles_Schwinger}), the violation of the reflection positivity can be easily checked by observing the non-positivity of $\int d^3\vec{\xi}~ S(\vec{\xi},\xi_4)$.

(D)
We can show the Lorentz covariance as follows in the use of holomorphy and Euclidean rotation symmetry.
First, the Schwinger function is invariant under Euclidean rotations.
Then, the analytically-continued Wightman function is invariant under infinitesimal Euclidean rotations, so is invariant under its complexified version, namely infinitesimal complex Lorentz transformations.
Therefore, the reconstructed Wightman function is invariant under the restricted Lorentz group in the limit of going to the Minkowski spacetime.
One can also explicitly check the Lorentz invariance of the expression (\ref{eq:simple_complex_poles_hol_Wightman}) by a contour deformation.

For the case with a single scalar field, the locality, or the spacelike commutativity [$W(\xi) = W(-\xi)$ for spacelike $\xi$], is an immediate consequence from the Lorentz invariance.
For general cases, the locality follows from the permutation symmetry of the Schwinger function and the complex Lorentz covariance of the holomorphic Wightman function.

So far, we have seen general properties of complex singularities (A) -- (D).
We can now answer the question ($\alpha$).
\begin{itemize}
    \item [($\alpha$)] It is possible to reconstruct the Wightman function, and the Wightman and OS axioms are summarized in Tables \ref{tab:Wightman} and \ref{tab:OS} in the presence of complex singularities.
\end{itemize}

\begin{table}[t]
    \centering
\begin{tabular}{ |l|c| } 
 \hline
 ~[W0] Temperedness & violated \xmark \\ 
 ~[W1] Poincar\'e  Symmetry & preserved \gcheckmark \\ 
 ~[W2] Spectral Condition & violated \xmark   \\
 ~[W3] Spacelike Commutativity & preserved \gcheckmark \\
 ~[W4] Positivity & violated \xmark \\
 ~[W5] Cluster property & irrelevant \\
 \hline
\end{tabular}
    \caption{Wightman axioms for Wightman functions in the Minkowski spacetime.}
    \label{tab:Wightman}
\end{table}

\begin{table}[t]
    \centering
\begin{tabular}{ |l|c| } 
 \hline
 ~[OS0] Temperedness & assumed \gcheckmark  \\ 
 ~[OS1] Euclidean Symmetry & assumed \gcheckmark  \\ 
 ~[OS2] Reflection Positivity & violated \xmark \\
 ~[OS3] Permutation Symmetry & assumed \gcheckmark  \\
 ~[OS4] Cluster property & irrelevant \\
 ~[OS0'] Laplace transform condition & violated {\footnotesize (but irrelevant)} \\
 \hline
\end{tabular}
    \caption{OS axioms for Schwinger functions in the Euclidean space.}
    \label{tab:OS}
\end{table}

Let us make some comments on the results.

\begin{itemize}
    \item The exponential growth of the Wightman function (\ref{eq:Wightman}) in the limits $\xi^0 \rightarrow \pm \infty$ has far-reaching consequences.
This strongly suggests the ill-definedness of the corresponding S-matrix elements.
The states causing complex singularities should be therefore excluded from the physical sector by some confinement mechanism.
Moreover, the time-ordered propagator cannot be Fourier-transformed because of this exponential growth.
Thus, the simple inverse Wick rotation in the momentum space $k_E^2 \rightarrow - k^2$ cannot be applied in the presence of complex singularities.
    \item Complex singularities are often discussed to be associated with non-locality in some literature since they cannot appear in the usual formalism of local QFTs. However, from (D) the compatibility with the spacelike commutativity, complex singularities themselves do not necessarily lead to non-locality.
\end{itemize}

At first glance, from the violation of the temperedness, spectral condition, and positivity, complex singularities seem to have no interpretation.
However, we argue that complex singularities can appear in indefinite-metric QFTs.

\section{Realization in quantum theory}

Next, we consider the second question ($\beta$).
Since complex singularities are supposed to appear in the gluon propagator in the Landau-gauge Yang-Mills theory, it is natural to consider indefinite-metric QFTs.
An important observation is that complex-energy spectra can appear in an indefinite-metric state space.
States with complex conjugate eigenvalues of a hermitian Hamiltonian can be realized by zero-norm pairs:
\begin{align*}
    (\ket{E}, \ket{E^*})~
    \begin{cases}
    H \ket{E} = E \ket{E}, ~~~ H \ket{E^*} = E^* \ket{E^*} \\
    \langle E|E \rangle  = \langle E^*|E^* \rangle = 0,~~ \langle E|E^* \rangle \neq 0
    \end{cases}
\end{align*}
If such a pair exists, it contributes to the Wightman function as,
\begin{align*}
\langle 0 | \phi(t) \phi(0) | 0 \rangle &\supset (\langle E^* |E \rangle)^{-1} e^{- i E t} \bra{0} \phi(0) \ket{E} \bra{E^*} \phi(0) \ket{0} \\
    & ~~~ + (\langle E|E^* \rangle)^{-1} e^{- i E^* t} \bra{0} \phi(0) \ket{E^*} \bra{E} \phi(0) \ket{0}.
\end{align*}
By preparing a pair $(\ket{E}, \ket{E^*})$ for each momentum $\vec{p}$, we can reproduce the Wightman function reconstructed from a pair of complex poles (\ref{eq:Wightman}). Since a complex singularity can be basically expressed by a sum of complex poles, we reach the conclusion \cite{Hayashi:2021jju, Hayashi:2021nnj}:
\begin{itemize}
    \item [($\beta$)] Complex singularities can be realized in indefinite-metric QFTs and correspond to pairs of zero-norm eigenstates of complex energies.
\end{itemize}

To obtain a physical theory from an indefinite-metric QFT, we need to construct a physical state space.
A promising way is to use the Kugo-Ojima quartet mechanism \cite{Kugo:1979gm} by the BRST symmetry.
If this mechanism works well\footnote{Note, however, that it is highly nontrivial to see whether or not a nilpotent BRST symmetry exists in the Landau gauge adopted in the numerical works.}, the pairs of complex-energy states should be in BRST quartets.
In this light, it can be said that complex singularities correspond to confined states.
We can also argue that the existence of complex singularities in a propagator of the gluon-ghost composite operator is a necessary condition for this scenario\footnote{Incidentally, the Bethe-Salpeter equation for the gluon-ghost bound state was discussed in {\cite{Alkofer:2011pe}} from a similar point of view.}  \cite{Hayashi:2021jju}.

\section{Conclusion}

We have examined the reconstruction of propagators and its consequences in the presence of complex singularities.
In conclusion, the existence of complex singularities does not rule out the possibility to reconstruct a local QFT (with an indefinite metric) although complex singularities are out of the standard formalism of QFT as shown in Tables \ref{tab:Wightman} and \ref{tab:OS}.

\section*{Acknowledgements}

\paragraph{Funding information}

Y.~H. is supported by JSPS Research Fellowship for Young Scientists Grant No.~20J20215, and K.-I.~K. is supported by Grant-in-Aid for Scientific Research, JSPS KAKENHI Grant (C) No.~19K03840.



\nolinenumbers


\begin{thebibliography}{99}

\bibitem{BT2019}
D. Binosi and R.-A. Tripolt, 
Phys. Lett. B \textbf{801}, 135171 (2020),
\doi{10.1016/j.physletb.2019.135171}.


\bibitem{Falcao:2020vyr}
A.~F.~Falc\~ao, O.~Oliveira and P.~J.~Silva,
Phys. Rev. D \textbf{102}, 114518 (2020),
\doi{10.1103/PhysRevD.102.114518}.


\bibitem{SFK12}
S. Strauss, C.S. Fischer, and C. Kellermann,
Phys. Rev. Lett. \textbf{109}, 252001 (2012),
\doi{10.1103/PhysRevLett.109.252001}.

\bibitem{Fischer-Huber}
C.~S.~Fischer and M.~Q.~Huber,
Phys. Rev. D \textbf{102}, 094005 (2020),
\doi{10.1103/PhysRevD.102.094005}.











\bibitem{Siringo16}
F. Siringo,
Nucl.Phys. \textbf{B907}, 572 (2016),
\doi{10.1016/j.nuclphysb.2016.04.028};
F. Siringo,
Phys. Rev. D \textbf{94}, 114036 (2016),
\doi{10.1103/PhysRevD.94.114036}.

\bibitem{HK2018}
Y. Hayashi and K.-I. Kondo, Phys. Rev. D \textbf{99}, 074001 (2019),
\doi{10.1103/PhysRevD.99.074001}.

\bibitem{Hayashi:2020few}
Y.~Hayashi and K.-I.~Kondo,
Phys. Rev. D \textbf{101}, 074044 (2020),
\doi{10.1103/PhysRevD.101.074044}.



\bibitem{Stingl}
M.~Stingl,
Phys. Rev. D \textbf{34}, 3863 (1986), \doi{10.1103/PhysRevD.34.3863};
U. H\"abel, R. K\"onning, H. G. Reusch, M. Stingl, and
S. Wigard, Z. Phys. A \textbf{336}, 423 (1990),
\doi{10.1007/BF01294116};
U. H\"abel, R. K\"onning, H. G. Reusch, M. Stingl, and
S. Wigard, Z. Phys. A \textbf{336}, 435 (1990),
\doi{10.1007/BF01294117};
M. Stingl, 
Z. Phys. A \textbf{353}, 423 (1996),
\doi{10.1007/BF01285154}.


%
\bibitem{Hayashi:2021nnj}
Y.~Hayashi and K.-I.~Kondo,
Phys. Rev. D \textbf{103}, L111504 (2021),
\doi{10.1103/PhysRevD.103.L111504}.

%
\bibitem{Hayashi:2021jju}
Y.~Hayashi and K.-I.~Kondo,
Phys. Rev. D \textbf{104}, 074024 (2021),
\doi{10.1103/PhysRevD.104.074024}.


\bibitem{OS73}
K.~Osterwalder and R.~Schrader,
Commun. Math. Phys. \textbf{31}, 83 (1973),
\doi{10.1007/BF01645738}.

\bibitem{OS75}
K.~Osterwalder and R.~Schrader,
Commun. Math. Phys. \textbf{42}, 281 (1975),
\doi{10.1007/BF01608978}.





\bibitem{Kugo:1979gm}
T.~Kugo and I.~Ojima,
Prog. Theor. Phys. Suppl. \textbf{66}, 1 (1979),
\doi{10.1143/PTPS.66.1}.


\bibitem{Alkofer:2011pe}
N.~Alkofer and R.~Alkofer,
Phys. Lett. B \textbf{702}, 158 (2011),
\doi{10.1016/j.physletb.2011.06.073}.

\end{thebibliography}
\end{document}